\documentclass[prl,superscriptaddress,twocolumn,showkeys,floatfix]{revtex4-1}

\usepackage{amsmath}
\usepackage{amssymb}
\usepackage{bm}         
\usepackage{comment}    
\usepackage[T1]{fontenc}
\usepackage{graphicx}   
\usepackage[
colorlinks=true,
citecolor=MidnightBlue,
linkcolor=BrickRed,
urlcolor=blue,
breaklinks=true,
]{hyperref}   			
\usepackage{mathtools}	
\usepackage{subfigure}	
\usepackage[
normalem
]{ulem}					
\usepackage[
usenames,
dvipsnames
]{xcolor}				
\usepackage{pgfplots}
\usepackage{mathrsfs}

\newcommand{\bmv}{\bm{v}}

\newcommand{\kb}{k_{\mathrm{b}}}

\newcommand{\SL}{\mbox{\textit{S{\footnotesize{\hspace{-0.5pt}\textit{L}}}}}}	
\newcommand{\FvK}{F\"{o}ppl--von K\'{a}rm\'{a}n}

\newcommand{\mk}{\mkern1mu}

\newcommand{\SLac}{{\SL_{\mkern1mu\mathrm{ac}}}}
\newcommand{\SLbar}{\mkern4.5mu\overline{\mkern-4.5mu\SL\mkern-1.0mu}\mkern1.0mu}

\newcommand{\Gammac}{\Gamma_{\! \mathrm{c}}}

\newcommand{\Gammao}{\Gamma_{\! 1}}
\newcommand{\Gammat}{\Gamma_{\! 2}}

\newcommand{\md}{{\mathrm{d}\mkern1mu}}

\newcommand{\ik}{\mathrm{i}\mkern1mu}
\newcommand{\si}{{(i)}}
\newcommand{\sr}{{(r)}}
\newcommand{\omegaz}{\omega^{}_0}
\newcommand{\omegai}{\omega^\si}
\newcommand{\omegar}{\omega^\sr}
\newcommand{\omegaiz}{\omega^\si_0}

\newcommand{\qz}{q^{}_0}
\newcommand{\qi}{q^\si}
\newcommand{\qr}{q^\sr}

\newcommand{\mf}{{\mathrm{f}}}
\newcommand{\omegaf}{\omega^{}_\mf}

\newcommand{\omegarf}{\omega^\sr_\mf}
\newcommand{\qf}{q^{}_\mf}

\newcommand{\qrf}{q^\sr_\mf}

\newcommand{\SLf}{\SL_{\mkern1mu\mf}}
\newcommand{\Vf}{V_\mf}

\newcommand{\sz}{{(0)}}
\newcommand{\lambdaz}{\lambda^{}_0}
\newcommand{\rz}{r^{}_0}
\newcommand{\rzt}{r_0^{\, 2}}
\newcommand{\Vz}{V^{}_0}

\newcommand{\rd}[1]{r_{\! ,_{#1}}^{}}
\newcommand{\rh}{r_{\mathrm{h}}^{}}

\newcommand{\tilr}{\tilde{r}}

\newcommand{\bfrac}[2]{{^{#1} \! / \! _{#2}}}

\definecolor{myCyan}{rgb}{0.0, 0.47, 0.75}
\colorlet{myBlue}{blue!70!green}
\colorlet{myOrange}{red!60!yellow}
\colorlet{myRed}{orange!40!purple}
\colorlet{myGreen}{green!70!blue}
\colorlet{myPurple}{red!60!blue}
\colorlet{myDarkCyan}{myPurple!15!myCyan}

\makeatletter
\def\@fnsymbol#1{
	\ensuremath{\ifcase#1\or
	*\or				
	\ddagger\or			
	\dagger\or			
	\mathsection\or		
	\mathparagraph\or	
	\|\or				
	**\or				
	\ddagger\ddagger\or	
	\dagger\dagger		
	\else\@ctrerr\fi}}
\makeatother

\begin{document}

%
%

\preprint{APS/123-QED}

\title{Absolute vs Convective Instabilities and Front Propagation in Lipid Membrane Tubes}

\author{Jo\"el Tchoufag}
\email{jtchoufa@berkeley.edu}
\affiliation{
    Department of Chemical \& Biomolecular Engineering, University of California, Berkeley, CA 94720
}

\author{Amaresh Sahu}
\email{amaresh.sahu@berkeley.edu}
\affiliation{
    Department of Chemical \& Biomolecular Engineering, University of California, Berkeley, CA 94720
}

\author{Kranthi K. Mandadapu}
\email{kranthi@berkeley.edu}
\affiliation{
    Department of Chemical \& Biomolecular Engineering, University of California, Berkeley, CA 94720
}
\affiliation{
    Chemical Sciences Division, Lawrence Berkeley National Laboratory, Berkeley, CA 94720
}

\date{\today}

%
%

\begin{abstract}

We analyze the stability of biological membrane tubes, with and without a base flow of lipids.
Membrane dynamics are completely specified by two dimensionless numbers: the well-known \FvK\ number $\Gamma$ and the recently introduced Scriven--Love number $\SL$, respectively quantifying the base tension and base flow speed.
For unstable tubes, the growth rate of a local perturbation depends only on $\Gamma$, whereas $\SL$ governs the absolute or convective nature of the instability.
Furthermore, nonlinear simulations of unstable tubes reveal an initially localized disturbance results in propagating fronts, which leave a thin atrophied tube in their wake.
Depending on the value of $\Gamma$, the thin tube is connected to the unperturbed regions via oscillatory or monotonic shape transitions---reminiscent of recent experimental observations on the retraction and atrophy of axons. We elucidate our findings through a weakly nonlinear analysis, which shows membrane dynamics may be approximated by a model of the class of extended Fisher--Kolmogorov equations.
Our study sheds light on the pattern selection mechanism in axonal shapes by recognizing the existence of two Lifshitz points, at which the front dynamics undergo steady-to-oscillatory bifurcations.

\end{abstract}
\keywords{lipid membrane tube; absolute/convective instability; front propagation}

%
%

\maketitle

%
%

%
%

Lipid membranes are ubiquitous in biology, and are often found in cylindrical configurations \cite{terasaki-jcb-1986, lippencott-science-2016, roux-pnas-2002, koster-pnas-2003}.
Moreover, membrane tubes can sustain an axial flow of lipids, as observed in growing axons \cite{dai-bpj-1995, dai-cell-1995} and \textit{in vitro} cylindrical tethers \cite{Dommersnes2005}.
When such tubes are under high tension, an infinitesimal local perturbation grows and invades the undisturbed region via propagating fronts.
For example, drug treatments disrupting the internal structure of axons show fronts propagate from the neuron's growth cone to its soma, in the direction of lipid flow \cite{datar-bpj-2019}.
In contrast, when a laser is aimed at a point on an axon, two fronts propagate outward in different directions \cite{datar-bpj-2019}.
An additional complexity is that different patterns can be selected in the front wakes, leading to distinct morphologies at long times.
\begin{figure}[!t]
	\centering
	\includegraphics[width=0.95\linewidth]{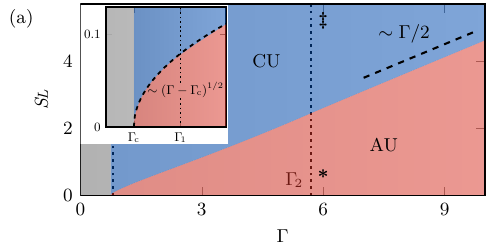}
	\hspace{15pt}
	\\[-2pt]
	\includegraphics[width=0.95\linewidth]{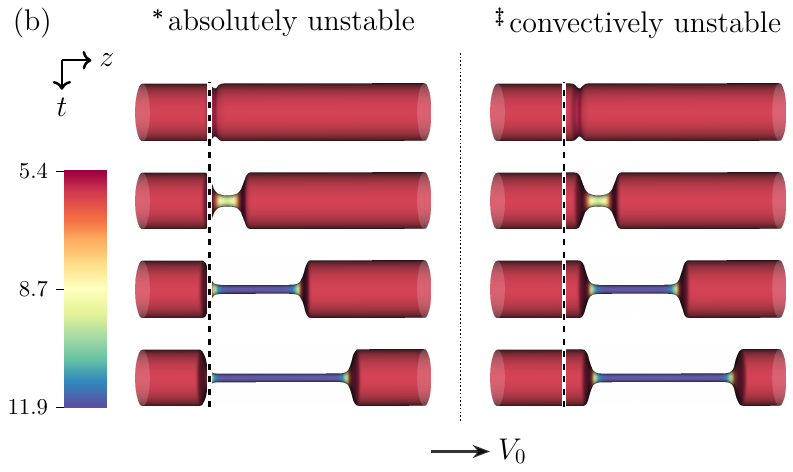}
	\vspace{-12pt}
	\caption{%
		(a) Stability diagram of lipid membrane tubes in the space of $\Gamma$ and $\SL$, showing stable (gray, $\Gamma < \Gammac = \bfrac{3}{4})$, absolutely unstable (AU, red), and convectively unstable (CU, blue) regimes.
		At the points marked `$\ast$' and `\ddag', nonlinear simulations (b) reveal propagating fronts.
		The initial perturbation is at the vertical dashed line, flow is in the $z$-direction, the color bar indicates surface tension, and snapshots are scaled $40 \times$ in the $z$-direction.
		\vspace{-20pt}
	}
	\label{fig:fig_abs_conv}
\end{figure}
For example, identical laser ablation experiments on axons result in a thin, atrophied tube---whose connection to the unperturbed regions appears pearled in some experiments and monotonic in others \cite{datar-bpj-2019}.
The latter are similar to our results in Fig.\ \ref{fig:fig_abs_conv}(b) and Movies S1--S3 in the Supplemental Material (SM)~\cite{supplemental}.
Despite many experimental findings, however, the physical mechanisms governing the front direction, the front speed, and the pattern selected in the front wakes remain poorly understood.%

In this Letter, we shed light on the aforementioned phenomena by incorporating two often overlooked features.
First, while most studies of cylindrical membranes assume the bulk fluid viscosity is the main dissipative cause \cite{nelson-prl-1995, granek-jp2f-1995, gurin-jetp-1996, goldstein-jp-1996, powers-prl-1997, boedec-jfm-2014, narsimhan-jfm-2015, al-izzi-prl-2018, bar-ziv-prl-1997, boedec-pre-2013}, here we recognize the intramembrane viscosity is the primary dissipative source \footnote{Reference \cite{al-izzi-prl-2020} considered azimuthal flows in tubes with an intramembrane viscosity; such flows are not investigated here.}.
Second, we incorporate a base flow of lipids in our description.
With these developments, we consider (for example) the lipid flow from growth cone to soma in a growing neuron and ask: when the axon is locally perturbed, is the disturbance only amplified downstream, or does it eventually invade the entire system?
This question is addressed by invoking the hydrodynamic concepts of \textit{absolute} and \textit{convective} instabilities \cite{bers-briggs-baps-1963, huerre-convective}, and analytically determining the critical base flow speed where the growth cone is on the verge of feeling a downstream perturbation at late times.
We conclude by investigating the dynamics of front propagation via the marginal stability criterion (MSC) \cite{Dee83, vanSaarloos89} and a weakly nonlinear analysis of membrane shape changes.
We find both atrophied and pearled morphologies can result from a local perturbation, and our front speed calculations qualitatively agree with past experiments.

\vspace{0.15cm}

%
%
\textbf{\textit{Membrane dynamics and dispersion relation.}}---%
We consider an unperturbed tube of radius $\rz$ along the $z$-axis.
The velocity field
$\bmv_\sz = \Vz \mk \bm{e}_z$,
for constant speed $\Vz$, captures the base lipid flow.
Moreover, a force balance \cite[Sec.\ I.4(a)]{supplemental} reveals the base surface tension is given by
$\lambdaz = p \, \rz + \kb / (4 \mk \rzt)$,
where $\kb$ is the bending modulus and $p$ is the jump in normal traction across the membrane
\footnote{%
	Some studies define
	$\kappa = \kb / 2$
	as the bending modulus.
	Dimensions:
	$[\kb] = \text{energy}$,
	$[\lambdaz] = \text{force}/\text{length}$,
	$[p] = \text{force}/\text{length}^2$.%
}%
.

Membrane dynamics have three sources of dissipation: the intermonolayer friction $b$ dominates at small length scales, the bulk viscosity $\mu$ dominates at large length scales, and the intramembrane viscosity $\zeta$ dominates across intermediate length scales
\footnote{%
	Dimensions:
	$[b] = \text{force} \cdot \text{time} / \text{length}^3$,
	$[\mu] = \text{force} \cdot \text{time} / \text{length}^2$,
	$[\zeta] = \text{force} \cdot \text{time} / \text{length}$.
	See also Refs.\ \cite{seifert-epl-1993, evans-yeung-cpl-1994, fournier-prl-2009, rahimi-arroyo-pre-2012, narsimhan-jfm-2015}.%
}%
.
For the systems of interest, dissipation from the bulk viscosity and intermonolayer friction are negligible \cite[Secs.\ VI.1, VI.2]{supplemental}, and membrane dynamics are governed by two dimensionless numbers.
The \FvK\ number $\Gamma$ compares tension forces to bending forces \cite{sahu-mandadapu-pre-2020, nelson-pre-2003}, and the recently introduced Scriven--Love number $\SL$ compares out-of-plane viscous forces to bending forces \cite{sahu-mandadapu-pre-2020}:
\begin{equation} \label{eq:eq_Gamma_SL_def}
	\Gamma
	\, = \, \dfrac{\lambdaz \, \rzt}{\kb}
	\qquad	\text{and}
	\qquad
	\SL
	\, = \, \dfrac{\zeta \, \Vz \, \rz}{\kb}
	~.
	\quad
\end{equation}
In past experiments with biological membrane tubes, we typically find
$\Gamma \in [\mk \bfrac{1}{4}, 3 \mk]$
and
$\SL \in [\mk 0, 0.04 \mk]$,
while in reconstituted cylinders
$\Gamma \in [\mk \bfrac{1}{4}, 7 \mk]$
and
$\SL \in [\mk 0, 1 \mk]$
(see SM \cite[Sec.\ V]{supplemental}).%

We now linearize the membrane equations about the chosen base state \cite[Sec.\ I.4]{supplemental}.
All quantities are non-dimensionalized with $\rz$, $\kb$, and $\zeta$---for which $\zeta \rzt / \kb$ is the fundamental time scale and $\kb / \rzt$ is the fundamental tension scale.
Lipid membrane tubes under tension, where
$\lambdaz \ge 0$,
are stable to all non-axisymmetric perturbations \cite[Sec.\ II.3]{supplemental}; \cite{narsimhan-jfm-2015}.
Consequently, only axisymmetric perturbations are considered here.
In this case, the perturbed equations of motion---in the absence of thermal noise
\footnote{%
    While thermal noise could be incorporated \cite{al-izzi-sm-2020}, such fluctuations do not contribute to the experiments of interest and are excluded here \cite[Sec.\ VI.3]{supplemental}.%
}---can be combined into a single equation governing the perturbed tube radius
$\tilr = \tilr(z, t)$
\footnote{
	When neglecting bulk dynamics, $p$ is assumed constant, and the tube volume is not conserved.
},
given by \cite[Sec.\ I.4(c)]{supplemental}
\vspace{-3pt}
\begin{equation} \label{eq:eq_evolution_equation_linear}
 	\tilr_{, t}
 	\, + \, \SL \, \tilr_{, z}
 	\, = \bigg(
 		\dfrac{\Gamma - \bfrac{3}{4}}{4}
 	\bigg) \, \tilr
 	\, + \bigg(
		\dfrac{\Gamma - \bfrac{1}{4}}{4}
 	\bigg) \, \tilr_{, z z}
 	\, - \, \dfrac{1}{8} \, \tilr_{, z z z z}
 	~.
\end{equation}

At this point, we decompose the perturbed radius $\tilr$ into normal modes of the form
$ \sim \! \exp [\mk \ik (q \mk z - \omega \mk t) \mk]$,
where $q$ is a dimensionless wavenumber and $\omega$ is a dimensionless frequency; the system is unstable when there exists a $q$ for which
$\omegai := \mathrm{Im} \mk \{\omega\} > 0$.
Substituting this decomposition into Eq.\ \eqref{eq:eq_evolution_equation_linear} leads to the dispersion relation
\vspace{-09pt}
\begin{equation} \label{eq:eq_axi_dispersion}
	\omega
	\, = \, \SL \, q
	\, + \, \dfrac{\ik}{4} \Big[
		\, \Gamma \big(
			1
			- q^2
		\big)
		\, + \, \dfrac{1}{4} \big(
			-3
			+ q^2
			- 2 \mk q^4
		\big) \,
	\Big]
	~.
\end{equation}
For normal modes with real wavenumber ($q \in \mathbb{R}$) and complex frequency ($\omega \in \mathbb{C}$), membrane tubes are unstable to a finite range of long wavelength perturbations when
$\Gamma > \Gammac := \bfrac{3}{4}$
\footnote{For bending modulus $\kappa = \kb/2$, the dimensionless critical tension is $3/2$.};
see SM \cite[Sec.\ II.3]{supplemental} and Refs.\ \cite{nelson-prl-1995, granek-jp2f-1995, gurin-jetp-1996, goldstein-jp-1996, powers-prl-1997, boedec-jfm-2014, narsimhan-jfm-2015, al-izzi-prl-2018}.
Though the growth rate of such sinusoidal perturbations is determined only by $\Gamma$, the real part of Eq.\ \eqref{eq:eq_axi_dispersion} reveals that a nonzero base flow ($\SL \ne 0$) leads to temporal oscillations in the membrane's response, with frequency proportional to base flow speed.
Additionally, an axial base flow confers a directionality to the membrane tube, biasing the spatial evolution of an initially local perturbation---bringing us to the concept of absolute and convective instabilities.

\vspace{0.05cm}

%
%
\textbf{\textit{Absolute vs.\ convective instabilities.}}---%
Consider an initially stable membrane tube subjected to a spatially localized perturbation, as in laser ablation \cite{bar-ziv-prl-1994} and local drug administration \cite{datar-bpj-2019} experiments.
Though such disturbances are local in space, they can globally alter $\lambdaz$ such that
$\Gamma > \Gammac$
along the tube \cite{goldstein-jp-1996}.
Here, we assume this tension change is instantaneous
\footnote{%
	In some complex cellular environments, tension may not spread rapidly from the source of a local perturbation~\cite{shi-cell-2018}.%
}.
Equations \eqref{eq:eq_evolution_equation_linear} and \eqref{eq:eq_axi_dispersion} then suggest that a base flow of lipids (captured by $\SL$) affects the directionality of the disturbance, while $\Gamma$ dictates how quickly the disturbance grows.
Figure \ref{fig:fig_abs_conv}(a) quantifies such observations, and shows how unstable tubes are either \textit{absolutely unstable} (AU) or \textit{convectively unstable} (CU) depending on $\Gamma$ and $\SL$.
The qualitative difference between AU and CU systems is highlighted in Fig.\ \ref{fig:fig_abs_conv}(b), where we consider what a stationary observer---for example, one stationed at the vertical dashed line---sees at long times.
In the AU case (Fig.\ \ref{fig:fig_abs_conv}(b), left), the base flow is sufficiently small and the initial perturbation invades the entire domain.
Consequently, our observer sees a deformed configuration at long times.
In the CU case (Fig.\ \ref{fig:fig_abs_conv}(b), right), however, the base flow is large enough to carry the perturbation downstream.
As a result, any stationary observer may see a transient growth of the instability, yet at long times will not see any effect from the perturbation---despite the disturbance continuing to grow as it is swept downstream.
In what follows, we determine when the system transitions from AU to CU, known as the \textit{absolute-to-convective transition}.

To determine the boundary between domains of AU and CU membrane tubes, we perform a spatiotemporal stability analysis \cite{bers-briggs-baps-1963, huerre-convective, charru}; see also the SM \cite[Sec.\ III]{supplemental}.
To this end, we consider complex wavenumbers
$q = \qr + \ik \qi \in \mathbb{C}$ 
and frequencies
$\omega = \omegar + \ik \omegai \in \mathbb{C}$ \cite{eggers-rpp-2008}.
We then search for the particular perturbations seen at long times in the laboratory frame---namely, those with zero group velocity, for which
$(\md \omega / \md q) \rvert_{\qz} = 0$
and
$\omegaz = \omega (\qz)$.
Here, $\qz$ and $\omegaz$ are the \textit{absolute wavenumber} and \textit{frequency}, and $(\qz, \mk \omegaz)$ is a saddle point of the system.
If $\omegaiz > 0$, a stationary observer sees the initial perturbation grow at long times and the system is AU (Fig.\ \ref{fig:fig_abs_conv}(b), left).
Otherwise, the tube is CU (Fig.\ \ref{fig:fig_abs_conv}(b), right).
Consequently, $\omegaiz = 0$
is a necessary condition at the boundary between AU and CU tubes, however it is not sufficient and the saddle point solutions could be spurious \cite{bers-briggs-baps-1963, kupfer-pf-1987, huerre-monkewitz-arfm-1990}.
We therefore supplement our search for the physical AU/CU transition with the pinching point method \cite{huerre-convective}, described in the SM \cite[Sec.\ III.2]{supplemental}.
We find that for every unstable
$\Gamma > \Gammac$,
there exists a critical Scriven--Love number $\SLac (\Gamma)$ corresponding to the AU/CU transition---shown as the boundary between AU  and CU domains in Fig.\ \ref{fig:fig_abs_conv}(a).
With the base flow speed and the calculated value of $\SLac (\Gamma)$, we predict the system's long-time response as seen by a stationary observer.
\begin{figure}[!t]
	\centering
	\includegraphics[height=0.42\linewidth]{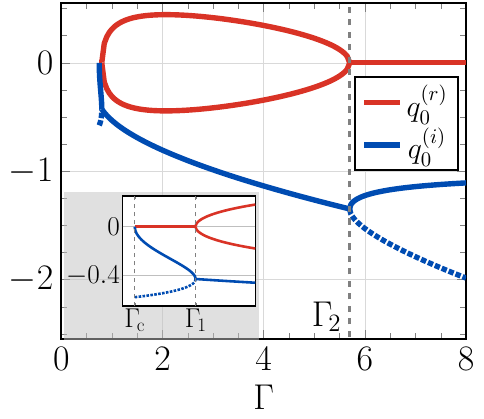}\hfill
	\includegraphics[height=0.42\linewidth]{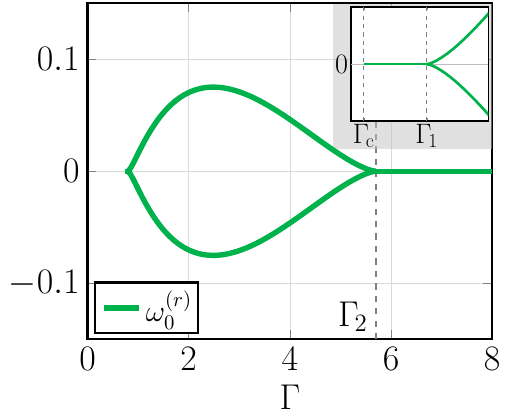}\\[-9pt]
	\caption{%
		Absolute wavenumber $\qz$ (left, units of $1/\rz$) and absolute frequency $\omegaz$ [right, units of $\kb / (\zeta \mk \rzt)$], with
		$\omegaiz = 0$.
		Solid (resp.\ dashed) lines correspond to relevant (resp.\ spurious) saddle points when determining the AU/CU boundary.
	}
	\label{fig:fig_bifurcation_q_omega}
	\vspace{-10pt}
\end{figure}

The analytical expression of $\SLac (\Gamma)$ is lengthy, and provided in Table 1 of the SM \cite{supplemental}.
Here we note two limiting cases:
$\SLac \approx \tfrac{\sqrt{2}}{4} ( \Gamma - \Gammac )^{1/2}$
when
$\Gamma \rightarrow \Gammac^+$,
and
$\SLac \approx \Gamma/2$
when
$\Gamma \rightarrow \infty$.
In calculating $\SLac (\Gamma)$, we also determine the absolute wavenumber $\qz (\Gamma)$ and frequency $\omegaz (\Gamma)$.
While $\SLac$ is a smoothly varying function of $\Gamma$, the saddle point $(\qz, \mk \omegaz)$ undergoes two bifurcations as $\Gamma$ is varied: one at
$\Gammao := \bfrac{13}{4} - \sqrt{6} \approx 0.8$
and another at
$\Gammat := \bfrac{13}{4} + \sqrt{6} \approx 5.7$,
as shown in Fig.\ \ref{fig:fig_bifurcation_q_omega} \cite[Sec.\ III.1]{supplemental}.
We subsequently discuss how the saddle point bifurcations affect the long-time response of a perturbed membrane tube.

\vspace{0.05cm}

%
%
\textbf{\textit{Front propagation.}}---%
When a membrane tube is locally perturbed,
the perturbed region invades the unstable, unperturbed region via a leading ($+$) and trailing ($-$) front [Fig.\ \ref{fig:fig_abs_conv}(b)], regardless of whether there is a base lipid flow.
The leading and trailing front velocities $\Vf^\pm$ of the linear theory are obtained with the MSC \cite{Dee83, vanSaarloos89}.
Namely, an observer traveling at the front speed would at long times see a system in its marginal state, where the growth rate is zero.
Applying the MSC, we find the dimensionless speed, wavenumber, and frequency of the front are respectively given by \cite[Sec.\ III.5]{supplemental}

~
\vspace{-11pt}
\begin{equation} \label{eq:eq_front_relations}
	\SLf^\pm
	\, = \, \SL
	\, \pm \, \SLac
	~,
	\quad
	\qf{^\pm}
	\, = \, \qz
	~,
	\quad
	\omegaf{^\pm}
	\, = \, \omegaz
	\, - \, \SLf^\pm \, \qf
	~.
\end{equation}
Equation \eqref{eq:eq_front_relations} predicts that $\SLac (\Gamma)$ is the front propagation speed when there is no base flow
($\SL = 0$).

\begin{figure}[!b]
	\vspace{-15pt}
	\centering
	\includegraphics[width=0.49\linewidth]{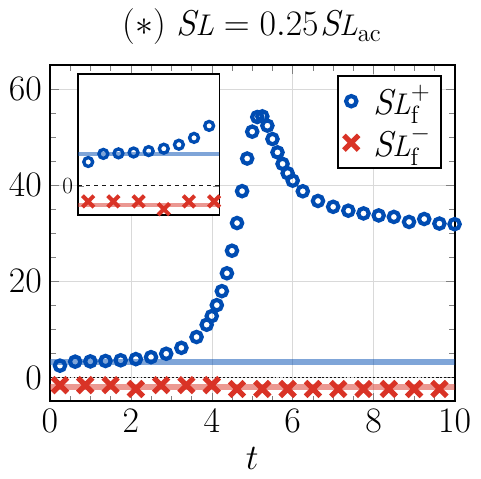}
	\includegraphics[width=0.49\linewidth]{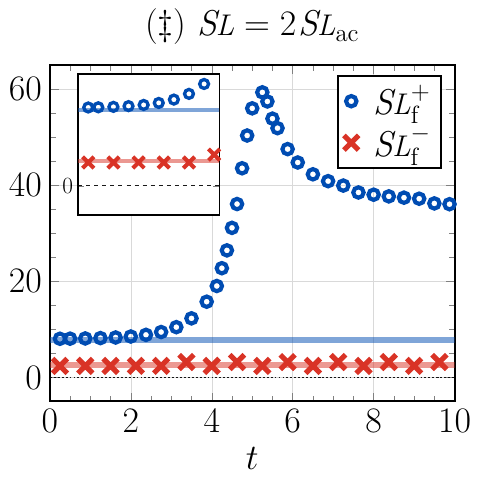}
	\\[-9pt]
	\caption{%
		Leading (blue) and trailing (red) front velocities over time, where $\Gamma = 6$ and $\SL/\SLac = $ 0.25 and 2 [cf.\ Fig.\ \ref{fig:fig_abs_conv}(b)].
		Symbols correspond to simulations and lines are predictions from the MSC \eqref{eq:eq_front_relations}$_1$.
		Insets show front speeds at early times.
	}
	\label{fig:fig_front_velocity}
\end{figure}

\begin{figure*}[!t]
	\centering
	\includegraphics[width=0.99\linewidth]{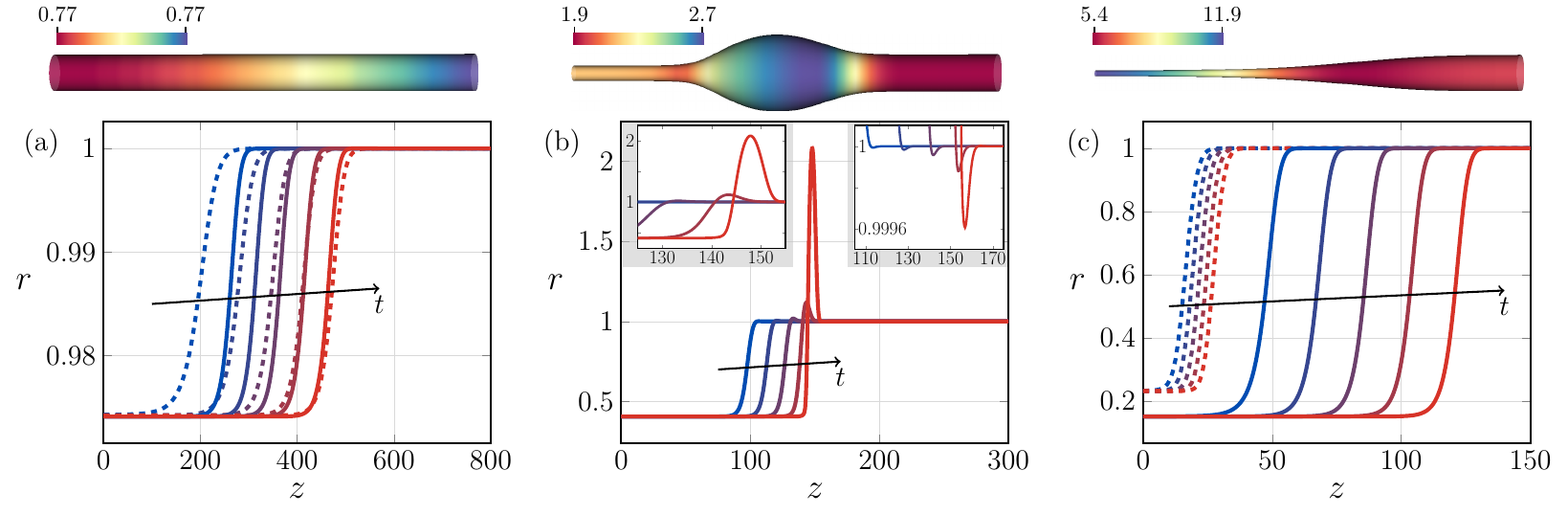}
	\\[-12pt]
	\caption{%
		Plots of propagating fronts over time, with no base flow
		($\SL = 0$)
		and
		(a) $\Gamma = 0.77 \in [\Gammac, \Gammao]$,
		(b) $\Gamma = 2    \in (\Gammao, \Gammat)$, and
		(c) $\Gamma = 6    \in [\Gammat, \infty) $.
		Solid (resp.\ dashed) lines are results from nonlinear simulations (resp.\ the EFK equation).
		Snapshots are separated by (a) $1000 \mk \tau$, (b) $10 \mk \tau$, and (c) $\tau$.
		A portion of the tube at the final snapshot is shown above each plot (color bars indicate surface tension); see also Movies S4--S7 \cite{supplemental}.
		When $\Gamma \in (\Gammao, \Gammat)$ as in (b),
		the front leaves pearls in its wake (left inset) and oscillates at the leading edge (right inset), in agreement with the linear theory; the outward bulges cause the EFK equation \eqref{eq:eq_evolution_weakly_nonlinear} to fail \cite[Sec.\ IV.2(b)]{supplemental}.
	}
	\label{fig:fig_nonlinear}
	\vspace{-10pt}
\end{figure*}

We now compare the linear predictions of Eq.\ \eqref{eq:eq_front_relations} to nonlinear simulations of an axisymmetric membrane tube subject to a local perturbation.
Our computations are based on those in Ref.\ \cite{omar-mandadapu-2019}; see also the SM \cite[Sec.\ IV.2]{supplemental}.
Figure \ref{fig:fig_front_velocity} compares the numerical front speeds to the linear prediction \eqref{eq:eq_front_relations}$_1$ for the two scenarios shown in Fig.\ \ref{fig:fig_abs_conv}(b).
It is well-known that a front can be either `pulled' at the leading edge or `pushed' by the growing nonlinearities behind the front, however one cannot in general anticipate which type of front will emerge from a local perturbation to invade a given nonlinear system \cite{vanSaarloos89, VanSaarloos09}.
Our numerical simulations reveal the trailing edge is a pulled front, where the front speed agrees with the MSC, while the leading edge is a pushed front traveling faster than the MSC prediction (Fig.\ \ref{fig:fig_front_velocity}).
In the latter case, the front speed agrees with the linear theory at early times when perturbations are small (Fig.\ \ref{fig:fig_front_velocity} insets).

Equation \eqref{eq:eq_front_relations} confirms the connection \cite{vanSaarloos1990, Chomaz2000} between the MSC and AU/CU transition: the saddle point bifurcations in Fig.\ \ref{fig:fig_bifurcation_q_omega} also represent transitions in the dynamics of the propagating fronts.
Specifically, when
$\Gamma \in [\Gammac, \Gammao] \cup [\Gammat, \infty)$,
then
$\qrf = 0$,
$\omegarf = 0$,
and the front evolves as a steadily traveling envelope---as confirmed by nonlinear simulations (Fig.\ \ref{fig:fig_nonlinear}(a),(c) and Movies S4, S7 \cite{supplemental}).
In contrast, when
$\Gamma \in (\Gammao, \Gammat)$,
$\qrf \ne 0$
and
$\omegarf \ne 0$%
---for which the front oscillates in both time and space.
As a result, a pattern is selected in the wake of the front, and a pearled morphology connects the thin and unperturbed cylindrical regions (Fig.\ \ref{fig:fig_nonlinear}(b), Movies S5, S6 \cite{supplemental}).
We thus find the \FvK\ number governs whether or not a pattern is selected as the front propagates.
Our results are consistent with recent observations of locally perturbed axons, where both monotonic and pearled fronts resulted \cite{datar-bpj-2019}.
The observed front speeds
($\Vf \sim 10^{-4}~\text{nm}/\mu\text{sec}$, $\SLf \sim 0.02$)
imply such axons lie close to the instability threshold $\Gammac$, with
$\Gamma \in [0.75, 1]$
(see SM \cite[Sec.\ V.5]{supplemental}).
Additionally, front speeds are the same order as the base flow speed in growing axons \cite{dai-bpj-1995, dai-cell-1995}, suggesting such systems lie close to the absolute-to-convective transition.

\vspace{0.05cm}

%
%
\textbf{\textit{Weakly nonlinear analysis.}}---%
To better understand how nonlinearities affect front dynamics, we develop a weakly nonlinear model of membrane shape evolution.
For a tube with dimensionless radius
$r(z, t) = 1 + \tilr(z, t)$,
we assume
(i) $\partial^j_z \tilr \cdot \partial^k_z \tilr$ is negligible when $j \ge 1$ and $k \ge 1$,
(ii) $\tilr_{, t} \cdot \partial^j_z \tilr$ is negligible when $j \ge 1$, and
(iii) $\tilr^j \cdot \partial^k_z \tilr \ll \tilr^j$ when $k \ge 1$.
With these assumptions, the radius evolves as \cite[Sec.\ IV.1]{supplemental}
\begin{equation} \label{eq:eq_evolution_weakly_nonlinear}
	\rd{T}
	\, + \, \SLbar \, \rd{Z}
	\, = \, \rd{Z Z}
	\, - \, \dfrac{1}{2} \, \dfrac{ (\Gamma - \Gammac) ~ }{ (\Gamma - \bfrac{1}{4})^2 } \, \rd{ Z Z Z Z}
	\, + \, f(r)
	~,
\end{equation}
with nonlinear forcing term
\begin{equation} \label{eq:eq_forcing_weakly_nonlinear}
	f(r)
	\, := \, (r - 1) 
	\, + \, \dfrac{\Gamma - \bfrac{1}{4}}{\Gamma - \Gammac} \, (r - 1)^2
	\, + \, \dfrac{\bfrac{1}{4}}{\Gamma - \Gammac} \, \dfrac{(r - 1)^2}{r}~.
    \vspace{-2pt}
\end{equation}
Hereafter,
$\gamma (\Gamma) := \tfrac{1}{2} (\Gamma - \Gammac)/(\Gamma - \bfrac{1}{4})^2$
is the $\partial^4 r / \partial Z^4$ coefficient in Eq.\ \eqref{eq:eq_evolution_weakly_nonlinear}.
Additionally, Eq.\ \eqref{eq:eq_evolution_weakly_nonlinear} is written with rescaled variables
$T = t \mk [ (\Gamma - \Gammac) / 4]$,
$Z = z \mk [ (\Gamma - \Gammac) / (\Gamma - \bfrac{1}{4})]^{1/2}$,
and
$\SLbar = 4 \mk \SL / [ (\Gamma - \Gammac) (\Gamma - \bfrac{1}{4}) ]^{1/2}$
to highlight its structure \footnote{The rescaled variables are only physically meaningful when $\Gamma > \Gammac$}.
Indeed, the evolution equation belongs to the family of extended Fisher--Kolmogorov (EFK) equations---known to possess several universal properties \cite{vanSaarloos89,Dee88,Cross93}.
For example, the EFK equation undergoes steady-to-oscillatory bifurcations in the front dynamics at the universal value of
$\gamma = \bfrac{1}{12}$ \cite{Dee88, Cross93}.
We confirm
$\gamma (\Gammao) = \gamma (\Gammat) = \bfrac{1}{12}$,
which justifies the observed transitions in front dynamics in Fig.~\ref{fig:fig_nonlinear}.
The values $\Gammao$ and $\Gammat$ are Lifshitz points where the $q^2$ term in the dispersion relation vanishes at the saddle point \cite{Zimmermann91}, and higher order gradients are required for an appropriate description of the system \cite[Sec.\ III.6]{supplemental}; see also Refs.\ \cite{Hornreich-1875,Cross93}.

Finally, we compare predictions from the EFK model \eqref{eq:eq_evolution_weakly_nonlinear} with simulations of the full nonlinear equations in Fig.\ \ref{fig:fig_nonlinear}.
Both sets of dynamics result in a thin, atrophied tube behind the front.
The atrophied `homogeneous radius' $\rh$ is calculated from the EFK equation \eqref{eq:eq_evolution_weakly_nonlinear} as
$\rh = 1 + (3 - 8 \mk \Gamma + \sqrt{ 16 \mk \Gamma - 3})/(8 \mk \Gamma - 2)$
\cite[Sec.\ IV.1]{supplemental}.
Thus, for all
$\Gamma > \Gammac$,
the evolution equation predicts
$0 < \rh < 1$.
Moreover,
$\rh \rightarrow 0$
in the limit
$\Gamma \rightarrow \infty$,
a result consistent with our previous findings \cite{sahu-mandadapu-jcp-2020}.
We also find that for
$\Gamma \notin (\Gammao, \Gammat)$,
the evolution equation is a good predictor of the front speed and final radius as
$\Gamma \rightarrow \Gammac^+$
[Fig.\ \ref{fig:fig_nonlinear}(a)],
while it only predicts $\rh$ at large $\Gamma$ [Fig.\ \ref{fig:fig_nonlinear}(c)].
On the other hand, when
$\Gamma \in (\Gammao, \Gammat)$,
the front oscillates and the quadratic forcing terms in $f(r)$ amplify outward perturbations
($r > 1$),
such that the EFK model predicts an unphysical, diverging radius at finite times \cite[Sec.\ IV.2(b)]{supplemental}.
Despite the quantitative shortcomings of the evolution equation, it still provides a connection to previous studies of nonlinear dynamics and pattern formation, and a qualitative understanding of front propagation.
Importantly, as the evolution equation \eqref{eq:eq_evolution_weakly_nonlinear} gives rise to pulled fronts and agrees with nonlinear simulations when $\Gamma\rightarrow\Gammac^+$, we expect the front velocity of an initially static tube to scale as $\SLf=\SLac \sim (\Gamma - \Gammac)^{1/2}$ near the instability threshold.
Though bulk dissipation is also important as
$\Gamma \rightarrow \Gammac^+$
\cite[Sec.\ VI.1]{supplemental}, our predicted power law scaling close to the instability threshold captures experimentally measured front velocities---a behavior that was previously unexplained (see Ref.\ \cite[Fig.\ 11]{bar-ziv-bpj-1998} and the SM \cite[Fig.\ 17]{supplemental}).

%
%
\textbf{\textit{Conclusions.}}---%
We analyzed the spatiotemporal stability of lipid membrane tubes, with and without a base flow.
Tubes were found to be either stable, AU, or CU depending on the values of $\Gamma$ and $\SL$, as shown in Fig.\ \ref{fig:fig_abs_conv}(a).
Our work was inspired by experimental observations of both pearled and atrophied membrane morphologies resulting from laser ablation experiments \cite{bar-ziv-prl-1997, datar-bpj-2019}.
By recalling the connection between the AU/CU transition and the MSC, we showed that unstable membrane tubes possess critical Lifshitz points at which propagating fronts bifurcate from a steady to an oscillating behavior, and vice versa (Fig.\ \ref{fig:fig_nonlinear}).
Moreover, our predictions for how $\Gamma$ selects a pattern are consistent with the aforementioned experiments.
While our study neglected the surrounding fluid viscosity, a previous analysis of front propagation in membrane tubes---which neglected the intramembrane viscosity yet included the bulk viscosity---always predicted a pearled configuration in unstable tubes \cite{goldstein-jp-1996}.
Thus, a natural extension of this work is to perform a spatiotemporal stability analysis which includes both intramembrane and bulk viscosities.

%
%

\vspace{0.05cm}

\begin{acknowledgments}
	\textbf{\textit{Acknowledgements.}}---%
	We thank Mr.\ Yannick Omar for helpful discussions regarding the axisymmetric membrane simulations, and Dr.\ Dimitrios Fraggedakis for insightful questions.
	J.T.\ acknowledges the support of U.T.\ Southwestern and U.C.\ Berkeley.
	A.S.\ acknowledges the support of the Computational Science Graduate Fellowship from the U.S.\ Dept.\ of Energy, as well as U.C.\ Berkeley.
	K.K.M.\ is supported by U.C.\ Berkeley.
\end{acknowledgments}

%
%

\bibliographystyle{apsrev4-1}
\bibliography{refs}

\end{document}